\documentclass[twocolumn, letter]{jpsj3}
\usepackage{txfonts}
\title{Cation Dependence of the Electronic States in Molecular Triangular Lattice System $\beta^\prime$-$X$[Pd(dmit)$_2$]$_2$: A First-Principles Study}  

\author{Takao \surname{Tsumuraya}$^{1, 2}$\thanks{E-mail: tsumu@riken.jp}
, Hitoshi \surname{Seo}$^{1, 3}$, 
Masahisa  \surname{Tsuchiizu}$^{4}$, 
Reizo  \surname{Kato}$^{1}$, and Tsuyoshi \surname{Miyazaki}$^{2}$}
\inst{$^{1}$RIKEN, Wako, Saitama  351-0198, Japan,\\
$^{2}$Computational Materials Science Unit, National Institute for Materials Science, Tsukuba 305-0047, Japan\\
$^{3}$JST-CREST, Wako, Saitama 351-0198, Japan\\ 
$^{4}$Department of Physics, Nagoya University, Nagoya 464-8602, Japan}
\abst{
The electronic structure of an isostructural series of molecular conductors, $\beta^\prime$-$X$[Pd(dmit)$_2$]$_2$, 
is systematically studied by a first-principles method based on the density-functional theory.
The calculated band structures are fitted to the tight-binding model based on Pd(dmit)$_2$ dimers on the triangular lattice. We find a systematic variation in the anisotropy of the transfer integrals along the three directions of the triangular lattice taking different values. The transfer integral along the face-to-face stacking direction of Pd(dmit)$_2$ dimers is always the largest. Around the quantum spin liquid, $X$ = EtMe$_3$Sb, the other two transfer integrals become comparable. 
We also report sensible differences in the distribution of wavefunctions near the Fermi level between the two dmit ligands of the Pd(dmit)$_2$ molecule. 
} 
\kword{first-principles calculations, density functional theory, molecular conductor, tight-binding model}

\begin{document}
\maketitle
Electronic and magnetic properties of a series of anion radical salts, $X$[Pd(dmit)$_2$]$_2$ have attracted much attention recently.~\cite{Kanoda_Kato_2011ARCMP}
Here, monovalent cations $X$ take Et$_{y}$Me$_{4-y}Z$, where $y$ = 0 -- 2, Et = C$_2$H$_5$, 
Me = CH$_3$, $Z$ = P, As, and Sb, and dmit denotes 1,3-dithiol-2-thione-4,5-dithiolate. 
Most of the salts are isostructural at room temperature (RT) with the space group $C2$/$c$ (called the $\beta'$-type structure).~\cite{Crystals12_Kato} 
These salts are paramagnetic Mott insulators at RT and exhibit a rich variety of ground states at low temperatures (LT) depending on the cation $X$.
Many of them show antiferromagnetic (AF) order~\cite{Tamura2002} as in typical Mott insulators, while others show interesting ground states, such as the valence bond solid (VBS) state~\cite{Tamura_EtMe3P06} observed in the EtMe$_3$P salt, the quantum spin-liquid (QSL) state in the EtMe$_3$Sb salt~\cite{Itou_dmit_PRB2008, M_Yamashita_dmit10, Yamashita2011}, 
and the nonmagnetic charge-ordered (CO) state in the Et$_2$Me$_2$Sb salt.~\cite{Nakao_JPSJ_Et2Me2Sb, CO-interTheory}
It was also reported that the EtMe$_3$P salt~\cite{Kato_EtMe3P06, EtMe3P_PRL99} and the salts showing the AF order become superconducting with the application of a hydrostatic or uniaxial pressure of less than 1~GPa.~\cite{Kato_Chem_Rev}  
To elucidate the origin of such a variety of ground states, it is necessary to clarify the cation dependence of the electronic structure. 

In this Letter, we report a systematic study on the electronic structure of 
$\beta^\prime$-$X$[Pd(dmit)$_2$]$_2$ by first-principles density-functional theory (DFT) calculations. 
The differences between the present results and the results of previous studies obtained using the tight-binding model with the H\"{u}ckel method (H\"{u}ckel+TB)~\cite{Kanoda_Kato_2011ARCMP, Kato_Chem_Rev, Kato_PSSB2012, Crystals12_Kato} and DFT calculations~\cite{Powell_PRL12, Nakamura} will be discussed. 
In addition, we present a detailed analysis of the character of Kohn-Sham orbitals (wavefunctions) near the Fermi level.
\begin{figure}[b]
\begin{center}
\includegraphics[width=1.0\linewidth]{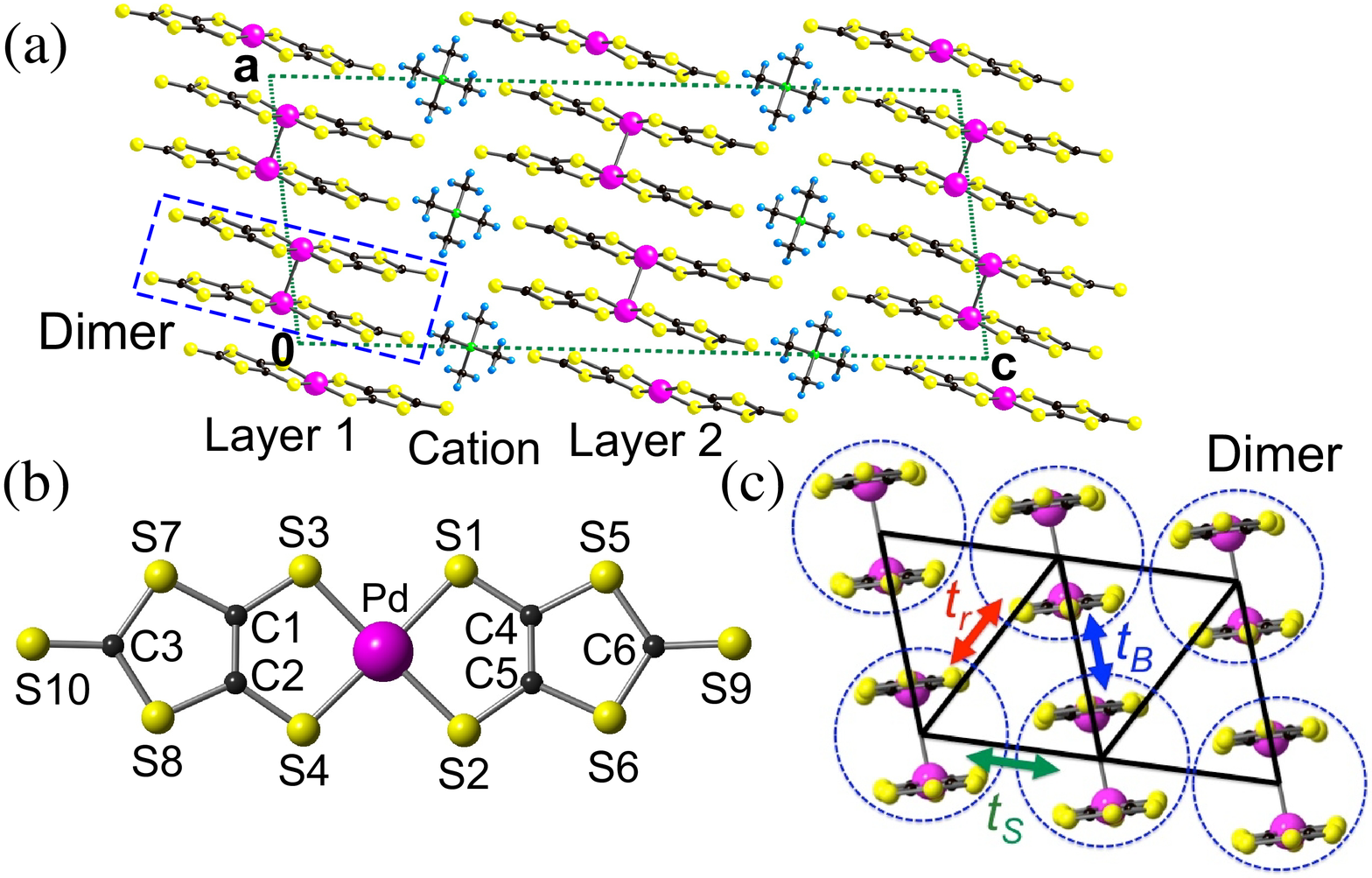}
\end{center}
\setlength\abovecaptionskip{0pt}
\caption{(Color online) (a) Crystal structure of $\beta^\prime$-Me$_4$P[Pd(dmit)$_2$]$_2$ ($C$2/$c$) at room temperature.~\cite{Crystals12_Kato} 
Dotted lines represent the conventional unit cell. 
(b) Molecular structure of Pd(dmit)$_2$. (c) Longitudinal view along the $c$-axis showing the $ab$-plane. The interdimer transfer integrals $t_B$, $t_S$ and $t_r$ are shown.}
\label{Structure}
\end{figure}

The first-principles DFT calculations were performed by all-electron full-potential linearized augmented plane wave 
(FLAPW) method~\cite{Wimmer1981, L_KA, Weinert}, which is one of the most reliable and accurate DFT methods.~\cite{EcutMT} 
The exchange-correlation functional was treated within the generalized gradient approximation (GGA) 
of the Perdew-Burke-Ernzerhof (PBE) functional.~\cite{GGA_PBE}. 
Non-spin-polarized calculations were performed using experimental crystal structures without geometrical optimization. 

Here, we work on nine members of $\beta^\prime$-type salts, which have the same space group $C$2/$c$ at RT.
Note that as for the EtMe$_3$P salt, the structure of a minor phase with $C$2/$c$ was used, whereas its main phase showing the VBS state has a different space group $P$2$_1$/$m$.~\cite{Tamura_EtMe3P06} 
As shown in Fig.~\ref{Structure}(a), $\beta^\prime$-$X$[Pd(dmit)$_2$]$_2$ has a layered structure along the $c$-axis; 
the anion and cation layers consist of Pd(dmit)$_2$ molecules and Et$_{y}$Me$_{4-y}Z$, respectively.~\cite{EtMe3Z}  
There are two equivalent anion layers in the unit cell, where Pd(dmit)$_2$ molecules [Fig.~\ref{Structure}(b)] stack face-to-face along the $[1 1 0]$ and $[1 {\bar 1} 0]$ directions with a strong dimerization (solid crossing column structure~\cite{Kato_Chem_Rev}). 
The Pd(dmit)$_2$ dimers form an anisotropic triangular arrangement in each anion layer as shown in Fig.~\ref{Structure}(c). 

\begin{figure}[tb]
\begin{center}
\includegraphics[width=1.0\linewidth]{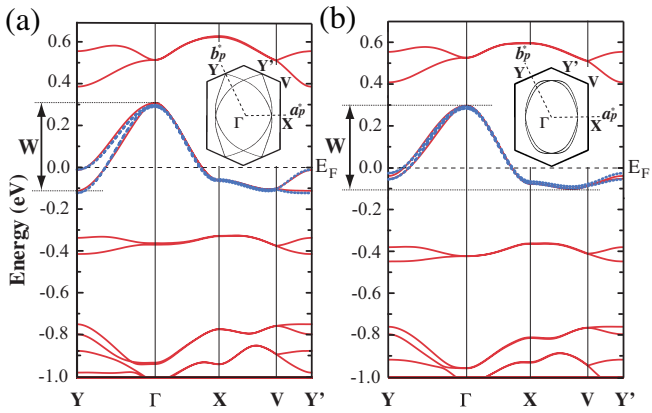}
\end{center}
\setlength\abovecaptionskip{0pt}
\caption{(Color online) Band structure and Fermi surface of $\beta'$-$X$[Pd(dmit)$_2$]$_2$ where $X$ = (a)  Me$_4$P and (b) EtMe$_3$Sb calculated by the first-principles DFT method.  
Dashed curves show the fitted results of the single-band TB model. The origin of the vertical axis with a dashed line shows the Fermi level. The present primitive unit cells ($a_p$ and $b_p$) and the labels of $k$-points are set to be the same as in the previous study.~\cite{Kato_Chem_Rev}} 
\label{Bands_Fermi}
\end{figure}

Figure~\ref{Bands_Fermi} shows the calculated band structures and Fermi surfaces of $\beta^\prime$-$X$[Pd(dmit)$_2$]$_2$, where $X$ = (a) Me$_4$P and (b) EtMe$_3$Sb. 
Consistent with the previous reports, the bands crossing the Fermi level are isolated from the other bands and are 1/2-filled.~\cite{Miyazaki99_dmit, Miyazaki03_dmit}
Previous studies suggest that these bands mainly originate from the antibonding pair of the highest occupied molecular orbitals (HOMO) of the two Pd(dmit)$_2$ molecules forming a dimer.~\cite{Kanoda_Kato_2011ARCMP, Kato_Chem_Rev, Canadell89, Canadell90_SScom, Miyazaki99_dmit} 
We will discuss this aspect and the character of the Kohn-Sham orbitals later.

\begin{figure}[b]
\begin{center}
\includegraphics[width=1.0\linewidth]{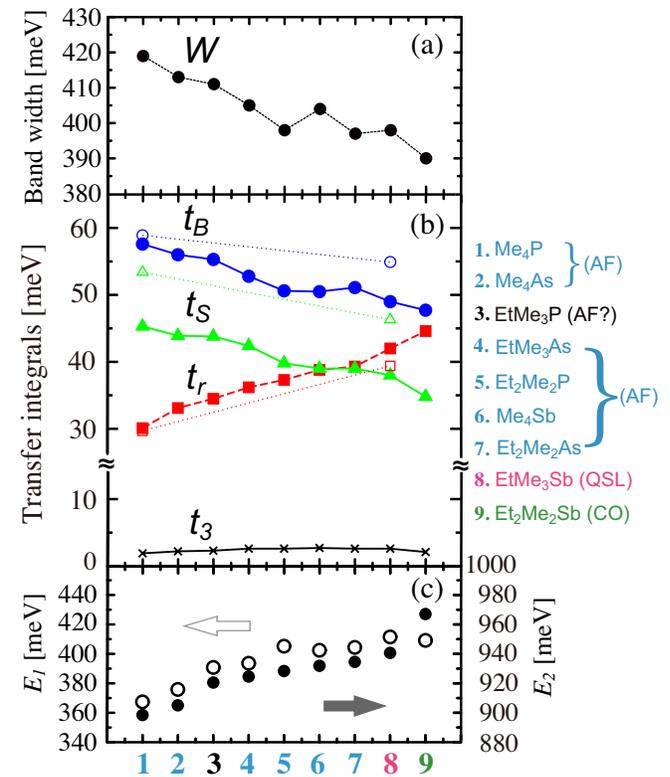}
\end{center}
\setlength\abovecaptionskip{0pt}
\caption{(Color online) Cation dependence of (a) bandwidth $W$, (b) interdimer transfer integrals, $t_B$, $t_S$, and $t_r$ and interlayer transfer integral $t_3$ shown by solid circles, solid triangles, and solid squares, and crosses, respectively. 
The corresponding open marks indicate the transfer integrals obtained using the low-temperature structures. 
(c) Open circles show $E_1$, the energy difference between the HOMO and LUMO states of an isolated neutral Pd(dmit)$_2$ dimer, and solid circles show $E_2$, the energy difference between the HOMO and LUMO+1 of the dimer, as shown in Fig.~\ref{LDOS}(a).}
\label{CationDep}
\end{figure}

Now, let us focus on the cation dependence. 
Figure~\ref{CationDep}(a) shows the cation dependence of the bandwidth $W$ of the bands crossing the Fermi level. 
The horizontal axis refers to the type of salt, i.e., $X$; we plot the salts according to the literature based on the H\"{u}ckel + TB scheme.~\cite{Kato_Chem_Rev, Tamura09, Kanoda_Kato_2011ARCMP, Crystals12_Kato}
The corresponding experimental ground states vary with decreasing N$\acute{\rm{e}}$el temperature to the QSL and CO states. 
Note that the magnetic structure of the $\beta^\prime$-type EtMe$_3$P salt at LT is unknown; thus, its order is determined by $W$.  
We find that $W$ monotonically decreases as the number of ethyl (Et) groups $y$ increases,
or the size of central atom $Z$ in terms of the covalent radius becomes larger, from P, As, to Sb. 
This tendency is consistent with the chemical pressure effect; larger cations (negative pressure) increase the volume of the unit cell and then, reducing $W$. 
It is discussed that $W$ can be considered as a control parameter of the electron correlation if the effective Coulomb parameter does not show a marked cation dependence.~\cite{Kato_Chem_Rev} 
On the other hand, it is reported that the anisotropy of the electronic structure is correlated to the observed ground states.~\cite{Kanoda_Kato_2011ARCMP} 
In this study, a characteristic cation dependence is seen in the anisotropy of the electronic structure shown in Fig.~\ref{Bands_Fermi}; the Fermi surfaces are composed of two equivalent cylinders associated with two anion layers. 
By changing the cation from Et$_2$Me$_2$Sb (No.9) to Me$_4$P (No.1), the cylinders are more distorted and less overlapped, indicating the electronic structure becomes more anisotropic. 

To evaluate the anisotropy quantitatively, we performed a tight-binding (TB) fit of the DFT bands and obtained the TB model using a single orbital of each dimer~\cite{Kanoda_Kato_2011ARCMP, Kato_Chem_Rev}  
with three interdimer transfer integrals, $t_B$, $t_S$, and $t_r$, in the Pd(dmit)$_2$ layer, as defined in Fig.~\ref{Structure}(c). 
An interlayer transfer integral across the cation layer, $t_3$, is also included. 
This simple TB model well reproduces the DFT bands in all the cases (only two cases are shown in Fig.~\ref{Bands_Fermi}). 
The obtained transfer integrals with different cations are plotted in Fig.~\ref{CationDep}(b). 
We find the following aspects: (1) $t_B$ is the largest. As the cation changes from left to right, 
(2) $t_S$ and $t_B$ decrease. (3) On the other hand, $t_r$ increases. 
The anisotropy represented by these transfer integrals well explains the cation dependence of the Fermi surfaces. 
Aspect (2) is consistent with the chemical pressure effect on $W$, whereas aspect (3) is peculiar. 
This may be explained by the unique distortions of the Pd(dmit)$_2$ molecule reported recently, which result in the shorter intermolecular distances between S atoms along the diagonal direction even though the volume increases.~\cite{Kato_PSSB2012, Crystals12_Kato}

For the materials on the left-hand side showing AF order, $t_S$ is greater than $t_r$. 
However, as a result of (2) and (3), the order of the two integrals is reversed in the two right most salts, which show the QSL or CO state.
Since the experimental structure at LT is available for the EtMe$_3$Sb (4.5~K~\cite{Kato_unpub}) and Me$_4$P salts (8~K~\cite{Kato_Me4P_LT}), we have also calculated their band structures and fitted the DFT bands to the same TB model~(both maintain $C$2/$c$).
Compared with the results obtained using the RT structure, $t_B$ and $t_S$ increase, whereas $t_r$ is almost the same. 
Consequently, the orders of the two transfer parameters $t_r$ and $t_S$ do not change, but
they become close in the case of the EtMe$_3$Sb salt.
Aspect (1) still holds even though the LT structure is used. 
Recently, Nakamura $et$ $al.$ ~\cite{Nakamura} have performed a DFT study on the EtMe$_3$Sb salt with the LT structure and evaluated the transfer integrals between the Wannier orbitals 
constructed from the Kohn-Sham orbitals of the bands crossing the Fermi level. 
Their calculated parameter values are in fairly good agreement with the present values derived from our TB fits.

Let us summarize the present results in Fig.~\ref{CationDep}(b), in relation to the experimentally observed properties. 
First, $t_B$, the transfer integral along the direction parallel to the face-to-face staking of Pd(dmit)$_2$ molecules (along $[1 1 0]$ and $[1 {\bar 1} 0]$) is the strongest. 
This implies a one-dimensional (1D) structure as a first approximation. 
The 1D chains are connected by $t_S$ and $t_r$. 
In the salts showing AF order, $t_S$ is larger than $t_r$; however, this difference decreases in the compounds with lower N$\acute{\rm{e}}$el temperatures.
Then, the two parameters become comparable around the two salts showing the competing ground states; the EtMe$_3$Sb salt shows the QSL state and the Et$_2$Me$_2$Sb salt shows the CO state.~\cite{Kato_PSSB2012} 
Such a tendency indicates a possible relevance of effective 1/2-filled electronic models or Heisenberg models where 1D chains are coupled via two frustrated interactions on an anisotropic triangular lattice.~\cite{Yunoki_Sorella_06, Ogata_JPSJ07}  

Next, the present results are compared with the previous results obtained by H\"{u}ckel+TB calculations using the same experimental structures.~\cite{Crystals12_Kato}  
Aspects (2) and (3) agree with the previoues TB parameters between the antibonding HOMO of the dimers, whereas the aspect (1) differs. Since $t_B$ and $t_S$ are nearly the same in each salt, ${t_r}$/${t}$, where $t=(t_B+t_S)$/2, was proposed as a key parameter to characterize the electronic structure.
Actually, the H\"{u}ckel+TB calculations for the LT structure of the EtMe$_3$Sb salt show a slightly larger value of $t_B$ than $t_S$: $t_B$ = 34.7, $t_S$ = 32.6, and $t_r$ = 26.6 meV. 
However, the difference is much smaller than the present results derived from the DFT calculations.

Recently, Scriven and Powell have also employed first-principles DFT calculations and performed a similar systematic study of $\beta'$-$X$[Pd(dmit)$_2$]$_2$ by introducing the same effective dimer model.~\cite{Powell_PRL12} 
They present five salts, except $X$ = Me$_4$As, Me$_4$Sb, EtMe$_3$P, and EtMe$_3$As from the nine salts we have calculated.  
However, some of their TB parameters are different from ours. 
Their band structure of the EtMe$_3$Sb salt is in good agreement with the present one, except that their position of the Fermi level is different from that in Fig.~\ref{Bands_Fermi}(b) and that of a previous study.~\cite{Nakamura, Miyazaki03_dmit} 
It is unclear why such discrepancies appear between their DFT-derived TB parameters and ours as well as those in ref. \citen{Nakamura}. 

\begin{figure}[b]
\begin{center}
\includegraphics[width=1.0\linewidth]{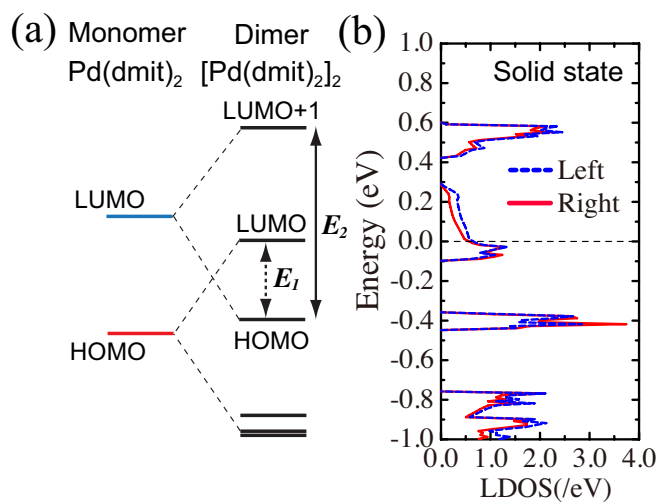}
\end{center}
\setlength\abovecaptionskip{0pt}
\caption{(Color online) (a) Calculated molecular orbital levels for an isolated monomer and its dimer using a supercell. The geometries are adapted from $\beta^\prime$-Et$_2$Me$_2$Sb[Pd(dmit)$_2$]$_2$ at room temperature. (b) Local density of states (LDOS) of $\beta^\prime$-Et$_2$Me$_2$Sb[Pd(dmit)$_2$]$_2$. Dashed curves indicate the LDOS from the left ligand, and solid curves, that from the right dmit ligand. 
A vertical dashed line represents the Fermi level. The LDOS of the left ligand includes partial DOS from C1--3, S3, S4, S7, S8, and S10, while that from the right includes from C4--6, S1, S2, S5, S6, and S9 atoms. [Fig.~\ref{Structure}(b)]}
\label{LDOS}
\end{figure}
Finally, we discuss the character of Kohn-Sham orbitals near the Fermi level.
It was suggested that the bands at the Fermi level and the bands located around --0.4 eV 
mainly come from the antibonding pair of HOMOs and the bonding pair of the lowest unoccupied molecular orbitals (LUMOs) of a Pd(dmit)$_2$ molecule forming the dimer, respectively.~\cite{Kanoda_Kato_2011ARCMP, Kato_Chem_Rev, Canadell89, Canadell90_SScom, Miyazaki99_dmit} 
We have calculated electron densities from the Kohn-Sham orbitals of these bands and found small but noticeable differences in the distribution of the orbitals between the left and right dmit ligands of the Pd(dmit)$_2$ unit in the crystal. This was not recognized in previous studies. This disproportionation within the molecule is observed also in the local density of states (LDOS) of the left and right dmit ligands, as shown in Fig.~\ref{LDOS}(b).
The contributions from the right are larger than those from the left at the bands located 
around --0.4 eV, while those from the left are larger than those from the right at the bands crossing the Fermi level.
In these calculations for the solid form, the disproportionation of the total charge density is small, almost negligible, compared with that of the charge density based on Kohn-Sham orbitals in each energy level.  
\begin{figure}[b]
\begin{center}
\includegraphics[width=1.0\linewidth]{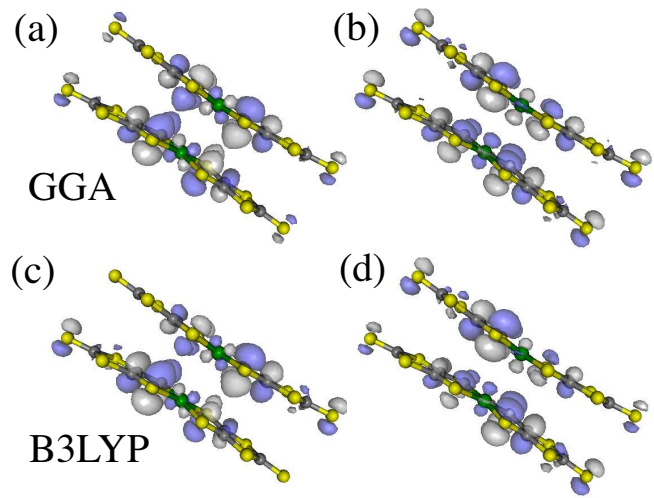}
\end{center}
\setlength\abovecaptionskip{0pt}
\caption{(Color online) Contour plots of molecular orbitals for an isolated dimer, [Pd(dmit)$_2$]$_2$$^{-}$,  calculated using the (a) HOMO and (b) singly occupied molecular orbital (SOMO) within GGA-PBE, and the (c) HOMO and (d) SOMO within B3LYP. The geometries are adopted from $\beta^\prime$-Et$_2$Me$_2$Sb[Pd(dmit)$_2$]$_2$ at room temperature.}
\label{dimerMO}
\end{figure}

A similar disproportionation is also seen in the corresponding molecular orbitals (MOs) of an isolated Pd(dmit)$_2$ dimer, but does not occur in an isolated monomer. 
The MOs and energy levels are calculated for one dimer/monomer in a supercell, whose atomic positions are extracted from the crystal structures of the nine salts. 
The cation dependence of the HOMO-LUMO gap of an isolated monomer is almost negligible ($\sim$0.61 eV within the GGA), but the energy differences between the dimer MOs ($E_1$ and $E_2$) show a systematic cation dependence [Figs.~\ref{CationDep}(c) and~\ref{LDOS}(a)].
This agrees well with the energy difference between the two bands of the solids, located at the Fermi level and around --0.4 eV. 
The two MOs of the isolated dimer also have noticeable differences in their amplitudes on the right and left dmit ligands, implying that the HOMO of the Pd(dmit)$_2$ monomer hybridizes with the LUMO of the other monomer within the dimer. This is possible since the HOMO-LUMO gap of the Pd(dmit)$_2$ monomer is relatively small~\cite{MO_ET} and the symmetry with respect to the left and right dmit ligands (the mirrors of $C_{2v}$~\cite{Rosa_dmit98}) is broken upon dimer formation in the solids. 
This molecular structure with the unique distortion is stable even after we optimize the internal coordinates using the DFT method.  

It is also noted that pure DFT methods, such as GGA, often underestimate such charge disproportionation. 
In fact, as shown in Fig.~\ref{dimerMO}, our calculated MOs of the isolated dimer by the B3LYP method~\cite{Beck3para98} using the Gaussian03 code~\cite{g03} show that the degree of disproportionation of the 
corresponding MOs is much larger than that by the GGA.~\cite{neutralMO} 
We expect that this degree of freedom within the molecule (coupled with the molecular distortion) will be confirmed by future experiments and may play important roles in the stability of various ground states, especially for the CO state~\cite{CO-interTheory}. 
A detailed investigation using the fragment decomposition scheme~\cite{Seo_JPSJ08_frag} is now in progress and will be reported elsewhere. 

In summary, we have performed a systematic study of $\beta^\prime$-$X$[Pd(dmit)$_2$]$_2$ by first-principles DFT calculations and clarified the cation ($X$) dependence of the electronic structure.
The dispersion of the DFT bands at the Fermi level is well reproduced by the effective TB model of the Pd(dmit)$_2$ dimer on the anisotropic triangular lattice. 
The transfer integrals show that the parameter along the face-to-face stacking direction of the Pd(dmit)$_2$ dimer ($[1 1 0]$ and $[1 {\bar 1} 0]$), $t_B$, is always larger than the other two parameters. 
We have also found that the other two parameters, $t_S$ and $t_r$, become comparable in the region where competing ground states (AF, QSL, and CO) are found experimentally. 
Furthermore, by analyzing the LDOS and Kohn-Sham orbitals of the solids, and the MOs of the isolated dimer, 
we revealed noticeable differences in charge distribution between the right and left dmit ligands of a Pd(dmit)$_2$ unit. 
This intramolecular degree of freedom may play important roles that clarify the various physical properties observed in the $\beta^\prime$-$X$[Pd(dmit)$_2$]$_2$ series. 

\begin{acknowledgment}
The authors would like to thank T. Itou for stimulating discussion. 
We acknowledge A. J. Freeman and S. Rhim for invaluable discussion about the FLAPW code.
This work was supported by Grant-in-Aid for Scientific Research (Nos. 20110003, 20110004, 24108511, and 24740232) from the MEXT. 
The computations in this work have been partly performed using the facilities of the RIKEN 
Integrated Cluster of Clusters (RICC), the numerical materials simulator at the National Institute for Materials Science (NIMS) Japan, and the National Energy Research Scientific Computing Center, which is supported by the Office of Science of the U.S. Department of Energy. 
\end{acknowledgment}
\bibliographystyle{jpsj}
\bibliography{./article_dmit_02}

\providecommand{\noopsort}[1]{}\providecommand{\singleletter}[1]{#1}%
\begin{thebibliography}{10}

\bibitem{Kanoda_Kato_2011ARCMP}
K.~Kanoda and R.~Kato: Annu. Rev. Condens. Matter Phys. {\bfseries 2} (2011)
  167.

\bibitem{Crystals12_Kato}
R.~Kato and H.~Cui: Crystals {\bfseries 2} (2012) 861.

\bibitem{Tamura2002}
M.~Tamura and R.~Kato: J. Phys.: Condens. Matter {\bfseries 14} (2002) L729.

\bibitem{Tamura_EtMe3P06}
M.~Tamura, A.~Nakao, and R.~Kato: J. Phys. Soc. Jpn. {\bfseries 75} (2006)
  093701.

\bibitem{Itou_dmit_PRB2008}
T.~Itou, A.~Oyamada, S.~Maegawa, M.~Tamura, and R.~Kato: Phys. Rev. B
  {\bfseries 77} (2008) 104413.

\bibitem{M_Yamashita_dmit10}
M.~Yamashita, N.~Nakata, Y.~Senshu, M.~Nagata, H.~M. Yamamoto, R.~Kato,
  T.~Shibauchi, and Y.~Matsuda: Science {\bfseries 328} (2010) 1246.

\bibitem{Yamashita2011}
S.~Yamashita, T.~Yamamoto, Y.~Nakazawa, M.~Tamura, and R.~Kato: Nat. Commun.
  {\bfseries 2} (2011) 275.

\bibitem{Nakao_JPSJ_Et2Me2Sb}
A.~Nakao and R.~Kato: J. Phys. Soc. Jpn. {\bfseries 74} (2005) 2754.

\bibitem{CO-interTheory}
M.~Tamura and R.~Kato: Chem. Phys. Lett. {\bfseries 387} (2004) 448 .

\bibitem{Kato_EtMe3P06}
R.~Kato, A.~Tajima, A.~Nakao, and M.~Tamura: J. Am. Chem. Soc. {\bfseries 128}
  (2006) 10016.

\bibitem{EtMe3P_PRL99}
Y.~Shimizu, H.~Akimoto, H.~Tsujii, A.~Tajima, and R.~Kato: Phys. Rev. Lett.
  {\bfseries 99} (2007) 256403.

\bibitem{Kato_Chem_Rev}
R.~Kato: Chem. Rev. {\bfseries 104} (2004) 5319.

\bibitem{Kato_PSSB2012}
R.~Kato, T.~Fukunaga, H.~M. Yamamoto, K.~Ueda, and H.~Cui: Phys. Status Solidi
  B {\bfseries 249} (2012) 999.

\bibitem{Powell_PRL12}
E.~P. Scriven and B.~J. Powell: Phys. Rev. Lett. {\bfseries 109} (2012) 097206.

\bibitem{Nakamura}
K.~Nakamura, Y.~Yoshimoto, and M.~Imada: Phys. Rev. B {\bfseries 86} (2012)
  205117.

\bibitem{Wimmer1981}
E.~Wimmer, H.~Krakauer, M.~Weinert, and A.~J. Freeman: Phys. Rev. B {\bfseries
  24} (1981) 864.

\bibitem{L_KA}
D.~D. Koellng and G.~O. Arbman: J. Phys. F: Metal Phys. {\bfseries 5} (1975)
  2041.

\bibitem{Weinert}
M.~Weinert: J. Math. Phys. {\bfseries 22} (1981) 2433.

\bibitem{EcutMT}
The cutoff energies of planewaves are 25 Ry for the LAPW basis functions and
  135 Ry for charge density. For Brillouin zone integrations,
  7$\times$7$\times$2 $k$-points are used. Muffin-tin sphere radii are set to
  be 1.06 \AA~for P, As, and Sb, 0.64~\AA~for C, 0.37~\AA~for H, 1.68~\AA~for
  S, and1.22~\AA~for Pd.

\bibitem{GGA_PBE}
J.~P. Perdew, K.~Burke, and M.~Ernzerhof: Phys. Rev. Lett. {\bfseries 77}
  (1996) 3865.

\bibitem{EtMe3Z}
In $C$2/$c$, cations are located on a two-fold axis, but EtMe$_3$Z does not
  have a two-fold symmetry. This is described by two possible orientations with
  an occupancy of 0.5 for each orientation by X-ray diffraction. In this study,
  EtMe$_3$$Z$ is replaced by Et$_2$Me$_2$Z to perform DFT calculations.

\bibitem{Miyazaki99_dmit}
T.~Miyazaki and T.~Ohno: Phys. Rev. B {\bfseries 59} (1999) R5269.

\bibitem{Miyazaki03_dmit}
T.~Miyazaki and T.~Ohno: Phys. Rev. B {\bfseries 68} (2003) 035116.

\bibitem{Canadell89}
E.~Canadell, I.~E.-I. Rachidi, S.~Ravy, J.~P. Pouget, L.~Brossard, and J.~P.
  Legros: J. Phys. France {\bfseries 50} (1989) 2967.

\bibitem{Canadell90_SScom}
E.~Canadell, S.~Ravy, J.~P. Pouget, and L.~Brossard: Solid State Commun.
  {\bfseries 75} (1990) 633.

\bibitem{Tamura09}
M.~Tamura and R.~Kato: Sci. Technol. Adv. Mater. {\bfseries 10} (2009) 024304.

\bibitem{Kato_unpub}
R. Kato $et$ $al.$: unpublished data.

\bibitem{Kato_Me4P_LT}
R.~Kato, Y.-L. Liu, Y.~Hosokoshi, S.~Aonuma, and H.~Sawa: Mol. Cryst. Liq.
  Cryst. {\bfseries 296} (1997) 217.

\bibitem{Yunoki_Sorella_06}
S. Yunoki and S. Sorella: Phys. Rev. B $\bf{74}$ (2006) 014408, and references
  therein.

\bibitem{Ogata_JPSJ07}
Y.~Hayashi and M.~Ogata: J. Phys. Soc. Jpn. {\bfseries 76} (2007) 053705.

\bibitem{MO_ET}
We have calculated the MOs of an isolated (BEDT-TTF)$_2$ dimer, whose structure
  is obtained from a solid, and found that the disproportionation of
  wavefunctions is negligibly small.

\bibitem{Rosa_dmit98}
A.~Rosa, G.~Ricciardi, and E.~J. Baerends: Inorg. Chem. {\bfseries 37} (1998)
  1368.

\bibitem{Beck3para98}
A.~D. Becke: J. Chem. Phys. {\bfseries 98} (1993) 5648.

\bibitem{g03}
Gaussian 03, www.gaussian.com.

\bibitem{neutralMO}
Their calculated MOs for an isolated neutral dimer, [Pd(dmit)$_2$]$_2$$^{0}$,
  show a similar charge disproportionation.

\bibitem{Seo_JPSJ08_frag}
H.~Seo, S.~Ishibashi, Y.~Okano, H.~Kobayashi, A.~Kobayashi, H.~Fukuyama, and
  K.~Terakura: J. Phys. Soc. Jpn. {\bfseries 77} (2008) 023714.

\end{thebibliography}
\end{document}